\documentclass[
reprint,
superscriptaddress,
amsmath,amssymb,
aps,
prb,
]{revtex4-2}

\usepackage{graphicx}% Include figure files
\usepackage{dcolumn}% Align table columns on decimal point
\usepackage{bm}% bold math
\usepackage{hyperref}% add hypertext capabilities
\usepackage{lineno}
\usepackage{orcidlink}

\hypersetup{
    colorlinks=true,
    allcolors=blue
}

\begin{document}

\title{Thermal conductivity of seifertite and pyrite-type SiO\textsubscript{2}: A comparative study}% 

\author{Doyoon Park\,\orcidlink{0009-0007-1224-6525}}
\email{doyoon.park@princeton.edu}
\affiliation{\mbox{Department of Mechanical and Aerospace Engineering, Princeton University, Princeton, New Jersey 08544, USA}}

\author{Yihang Peng\,\orcidlink{0000-0003-4404-8973}}
\affiliation{Department of Geosciences, Princeton University, Princeton, New Jersey 08544, USA}

\author{Jie Deng\,\orcidlink{0000-0001-5441-2797}}
\email{jie.deng@princeton.edu}
\affiliation{Department of Geosciences, Princeton University, Princeton, New Jersey 08544, USA}

\date{\today}

\begin{abstract}
Thermal conductivity is a fundamental material property that plays a crucial role in understanding the dynamics and evolution of planetary interiors. Despite its importance, the thermal conductivity of seifertite and pyrite-type SiO$_2$ remains unknown. Here, we calculate the lattice thermal conductivities of seifertite and pyrite-type SiO$_2$ using the Green-Kubo method based on molecular dynamics (MD) simulations driven by two machine learning potentials (MLPs) constructed from the SCAN and PBEsol exchange-correlation functionals, with \textit{ab initio}-level accuracy. To demonstrate our methodology, we also compute thermal conductivities using the phonon quasiparticle approach for comparison. Overall, the Green-Kubo method predicts up to 119 \% higher thermal conductivity with a temperature dependence close to $T^{-1}$, as it fully captures diffusion-like phonons at high temperatures that are missed by the phonon quasiparticle approach. The 19 \% reduction in thermal conductivity across the phase transition from seifertite to the pyrite-type phase suggests the potential formation of a thermally insulating layer in the mantle of super-Earths.

% \begin{description}
% % \item[Usage]
% % Secondary publications and information retrieval purposes.
% \end{description}
\end{abstract}

%Use showkeys class option if keyword display desired
\maketitle

\section{\label{sec:level1}Introduction}
Thermal conductivity plays a key role across many fields of science and technology, as it is directly linked to heat dissipation, thermal shielding, and energy efficiency \cite{cahill2003}. In planetary science, thermal conductivity is central to understanding the thermal evolution of planets. The early stages of rocky planets are often thought to be molten or partially molten due to impacts during planet formation \cite{elkins_tanton2012}, and thermal conductivity controls their subsequent cooling and solidification because heat removal from the interior is regulated by heat transport through thermal boundary layers \cite{buffett2002,naliboff2006}.

Silica (SiO$_2$) is one of the most abundant materials in rocky planets \cite{umemoto2017}. Despite their importance for understanding the dynamics of Earth's and exoplanetary interiors, the thermal conductivities of several high-pressure SiO$_2$ polymorphs remain unknown. Here, we focus on seifertite ($\alpha$-PbO$_2$-type, $Pbcn$) and pyrite-type ($Pa\bar{3}$) SiO$_2$, which are expected to be stable above $\sim$90 GPa and $\sim$240 GPa, respectively \cite{deng2026, park2026}. Experimentally measuring the thermal conductivity of such high-pressure polymorphs at high temperatures is challenging and subject to significant uncertainties \cite{kang2006,hasegawa2012}, and therefore theoretical calculations serve as effective alternatives.

The calculation of lattice thermal conductivity typically relies on three distinct methods: Boltzmann's kinetic approach \cite{klemens1958}, Fourier's law using non-equilibrium molecular dynamics (MD) \cite{muellerplathe1997}, and the Green-Kubo formalism \cite{green1954,kubo1957} coupled with equilibrium MD. The phonon Boltzmann transport equation (BTE) is generally formulated within an anharmonic lattice-dynamical framework for crystalline solids, where heat transport is described in terms of well-defined phonon quasiparticles vibrating about equilibrium atomic positions \cite{tang2010,dekura2019,zhang2021}. Therefore, its application is limited to temperatures well below melting or other structural phase transitions \cite{barbalinardo2020}. In contrast, the Green-Kubo method is free from such limitations and can be applied to any phase of matter, making it one of the most widely used approaches for determining thermal conductivity using classical interatomic potentials \cite{sellan2010}. However, classical potentials are likely to yield unreliable results under high-pressure and high-temperature conditions \cite{stackhouse2010}, because thermal conductivity is highly sensitive to the accuracy of the interatomic potential, and classical potentials often fail to capture material behavior under such extreme conditions. Furthermore, combining \textit{ab initio} calculations with the Green-Kubo method was considered challenging because it is difficult to decompose the total energy into individual atomic contributions. Therefore, non-equilibrium methods has also been used as an alternative in \textit{ab initio} calculations of thermal conductivity \cite{stackhouse2010}. However, compared with Green-Kubo method, it often requires larger simulation cells and long simulation times to establish a converged steady-state temperature gradient and heat flux \cite{sellan2010,schelling2002}. Although developments enabling the \textit{ab initio} Green-Kubo method have been reported \cite{marcolongo2016,carbogno2017}, the computational cost remains a substantial limitation.

Machine learning potentials (MLPs) provide \textit{ab initio}-level accuracy at significantly reduced computational cost, enabling simulations of large systems over long timescales \cite{behler2007,zhang2018}. Moreover, when using MLPs, the total energy of the system is naturally decomposed into individual atomic contributions, making them particularly well suited for the Green-Kubo method \cite{deng2021}. Therefore, the combination of MLPs and the Green–Kubo method enables the calculation of thermal conductivity with near-\textit{ab initio} accuracy at substantially reduced computational cost.

In this study, we report the lattice thermal conductivity of seifertite and pyrite-type SiO$_2$ using the Green-Kubo method with MD simulations powered by MLPs that accurately capture interatomic interactions based on the PBEsol \cite{perdew2008} and SCAN \cite{sun2015} exchange-correlation (XC) functionals. As a comparison and validation of our results, we also present the thermal conductivity calculated using the phonon quasiparticle method \cite{zhang2019,zhang2021}, which is based on the BTE and relaxtion-time approximation (RTA).  Although this phonon quasiparticle method accounts for temperature-dependent anharmonic phonon dispersions by combining density-functional perturbation theory (DFPT) \cite{baroni2001} and MD, it significantly underestimates the thermal conductivity compared to the Green-Kubo method at high temperatures. Using the Green-Kubo modal analysis (GKMA) \cite{lv2016}, which enables the modal decomposition of thermal conductivity within the Green-Kubo formalism, we find that the phonon quasiparticle approach has limitations in capturing the contribution of diffusion-like high-frequency phonons at high temperatures. Free from these constraints, the Green-Kubo method combined with MLPs captures additional transport channels not represented in the quasiparticle framework, even at extreme temperatures.

\section{Method}
\subsection{Machine learning potential}
An MLP is a neural network represenation of interatomic potential energy surface. In this work, we employ two MLPs developed by Park \textit{et al.} \cite{park2026}, based on the DeePMD-kit framework \cite{zhang2018,zeng2025} with an iterative training scheme \cite{deng2023}. These MLPs accurately capture interatomic interactions using the generalized gradient approximation (GGA) PBEsol and meta-GGA SCAN XC functionals for seifertite and pyrite-type SiO$_2$ over a broad range of pressures (100–400 GPa) and temperatures (1000–10000 K).

\subsection{Molecular dynamics simulations}
To generate equilibrium MD trajectories for the Green-Kubo method, we perform simulations using LAMMPS \cite{plimpton1995,thompson2022} interfaced with DeePMD-kit \cite{zeng2025}. We test a range of system sizes to examine the finite-size effects (Section~\ref{sec:convergence}). The systems are first relaxed using isothermal-isobaric (NPT) and canonical (NVT) ensemble simulations for 30 ps with the Nosé-Hoover thermostat \cite{hoover1985}. For seifertite, we consider three temperatures (2000, 3000, and 4000 K) and three pressures (140, 200, and 250 GPa). For the pyrite-type phase, we consider higher temperatures (3000, 4000, and 5000 K) and pressures (200, 250, and 400 GPa). After equilibration, the resulting structures are used as the initial configurations for microcanonical (NVE) ensemble simulations of 2 ns duration, from which the thermal conductivity is calculated. A timestep of 0.5 fs was used for the NVE simulations to minimize energy drift, while a 1 fs timestep was used for all other equilibration steps. To reduce statistical uncertainty, five independent MD simulation runs are performed for each temperature and pressure condition using different initial velocity distributions.

\subsection{Green-Kubo method}
According to the Green-Kubo formalism \cite{kubo1957}, the lattice thermal conductivity of a system is given as
\begin{equation}
\label{eq:1}
\kappa = \frac{1}{3 V k_B T^2} \int_{0}^{\infty} \langle \mathbf{J}(t_0+t) \cdot \mathbf{J}(t_0) \rangle \, dt,
\end{equation}
where $V$ is the volume of the cell, $k_B$ is the Boltzmann constant, $T$ is temperature, $t$ is the correlation time, $\langle...\rangle$ denotes average over time origins $t_0$, and $\mathbf{J}$ is the heat current which is defined as 
\begin{equation}
\label{eq:2}
\mathbf{J}
= \sum_{i=1}^{N} \dot{\mathbf{x}}_{i} \left( E_i + \frac{3}{2} k_B T \right)
- \sum_{\alpha=1}^{N_s} h_\alpha \sum_{i=1}^{N_\alpha} \dot{\mathbf{x}}_{\alpha i}
- \sum_{i=1}^{N} \mathbf{S}_i \cdot \dot{\mathbf{x}}_{i}
\end{equation}
where $N$ is the total number of atoms in the system, $\dot{\mathbf{x}}_{i}$ is the velocity of atom $i$, $E_i$ is the contributions of atom $i$ to the potential energy, $\mathbf{S}_i$ is the stress tensor of atom $i$, $N_s$ is the number of speices in the system, and $N_\alpha$ is the number of atoms of species $\alpha$. The partial enthalpy of species $\alpha$ is given as \cite{babaei2012,french2019}
\begin{equation}
h_\alpha = \frac{5}{2} k_B T
+ \frac{\sum_{i}^{N_\alpha}
\left(
E_i + \frac{1}{6} \sum_{\substack{j \neq i}}^{N} \mathbf{F}_{ij} \cdot \mathbf{r}_{ij}
\right)}{N_\alpha},
\end{equation}
where $\mathbf{F}_{ij}$ is the force acting on atom $i$ due to its interaction with atom $j$ and $\mathbf{r}_{ij}$ is the vector connecting the positions of atoms $i$ and $j$. This enthalpy term is crucial for thermal conductivity calculations in multicomponent liquid systems. However, by comparing the convergence of the heat current auto-correlation function
\begin{equation}
C(t) = \frac{1}{3 V k_B T^2}\langle \mathbf{J}(t_0+t) \cdot \mathbf{J}(t_0) \rangle,
\end{equation}
and the thermal conductivity (see Fig.~S1 of the Supplemental Material \cite{SM}), we find that it is also important to include the enthalpy term for SiO$_2$ seifertite and pyrite-type phases at high temperatures, even though they are relatively simple crystalline solids.

We use the centroid method, as implemented in LAMMPS, to calculate the per-atom stress tensor $\mathbf{S}_i$, because the non-centroid method is an approximation to the centroid method and can introduce errors of up to nearly 20 \% in the total heat flux, depending on the interatomic potential \cite{boone2019,surblys2019}. Further details of the centroid method are provided in Surblys et al. \cite{surblys2021}.

\subsection{Green-Kubo modal analysis}
The GKMA develped by Lv and Henry \cite{lv2016} enables calculation of modal contribution of thermal conductivity. By projecting the MD trajectory onto the eigenvecetors, we can write normal mode coordinates of position $X_n(t)$ and velocity $\dot{X}_n(t)$ as 

\begin{equation}
\label{eq:5}
X_n(t) = \sum_{i=1}^{N} \sqrt{m_i} \, \mathbf{e}_{i,n}^* \cdot \mathbf{x}_{i}(t),
\end{equation}

\begin{equation}
\dot{X}_n(t) = \sum_{i=1}^{N} \sqrt{m_i} \, \mathbf{e}_{i,n}^* \cdot \dot{\mathbf{x}}_{i}(t),
\end{equation}

where $m_i$ is mass of atom $i$, $\mathbf{e}_{i,n}$ is eigenvector for atom $i$ in mode $n$, and $\mathbf{x}_{i}$ and $\dot{\mathbf{x}}_{i}$ are displacement and velocity vector of atom $i$ respectively. By taking the reverse transformation, $\mathbf{x}_{i}$ and $\dot{\mathbf{x}}_{i}$ can be expressed as
\begin{equation}
\mathbf{x}_i(t) = \sum_{n=1}^{3N} \mathbf{x}_{i,n}(t) = \frac{1}{\sqrt{m_i}} \sum_{n=1}^{3N} \mathbf{e}_{i,n} X_n(t),
\end{equation}
\begin{equation}
\label{eq:8}
\dot{\mathbf{x}}_i(t) = \sum_{n=1}^{3N} \dot{\mathbf{x}}_{i,n}(t) = \frac{1}{\sqrt{m_i}} \sum_{n=1}^{3N} \mathbf{e}_{i,n} \dot{X}_n(t),
\end{equation}
using the normal mode coordinates. The formulation by Lv and Henry \cite{lv2016,seyf2019} used the heat current definition proposed by Hardy \cite{hardy1963}. However, because it is important to subtract the enthalpy term even for solid phases at high temperatures (Fig.~S1 \cite{SM}), we use Eq.~\eqref{eq:2} for heat current definition. Substituting the velocities in Eq.~\eqref{eq:2} with the normal mode coordinate expression shown in Eq.~\eqref{eq:8}, we obtain the heat current contribution for individual modes as
\begin{widetext}
\begin{equation}
\label{eq:9}
\mathbf{J}_n(t) = \sum_{i=1}^{N} \left(\frac{1}{\sqrt{m_i}}\mathbf{e}_{i,n}\dot{X}_n(t)\right) \left(E_i+\frac{3}{2}k_B T\right)
- \sum_{\alpha=1}^{N_s} h_\alpha \sum_{i=1}^{N_{\alpha}} \left(\frac{1}{\sqrt{m_i}}\mathbf{e}_{i,n}\dot{X}_n(t)\right)
- \sum_{i=1}^{N} \mathbf{S}_i\cdot \left(\frac{1}{\sqrt{m_i}}\mathbf{e}_{i,n}\dot{X}_n(t)\right),
\end{equation}
\end{widetext}
and the total heat current is given as $\mathbf{J} = \sum_{n=1}^{3N} \mathbf{J}_{n}(t)$. Thus, the modal contribution to the thermal conductivity is given by
\begin{equation}
\kappa_{n} = \frac{1}{3 V k_B T^2} \int_{0}^{\infty} \langle \mathbf{J}_n(t_0+t) \cdot \mathbf{J}(t_0) \rangle dt,
\end{equation}
and the total thermal conductivity can be calculated as
\begin{equation}
\kappa = \sum_{n=1}^{3N} \kappa_{n} = \frac{1}{3 V k_B T^2} \sum_{n=1}^{3N} \int_{0}^{\infty} \langle \mathbf{J}_n(t_0+t) \cdot \mathbf{J}(t_0) \rangle dt,
\end{equation}
which gives exactly the same result as Eq.~\eqref{eq:1} when using the same MD trajectory.
We can also write this as
\begin{equation}
\label{eq:13}
\kappa = \frac{1}{3 V k_B T^2} \int_{0}^{\infty} \big\langle \sum_{n=1}^{3N}\mathbf{J}_n(t_0+t) \cdot \sum_{n'=1}^{3N}\mathbf{J}_{n'}(t_0) \big\rangle dt,
\end{equation}
to emphasize its capability to analyze mode-mode correlations. The sum of the off-diagonal components ($n \ne n'$) in Eq.~\eqref{eq:13} corresponds to the contribution to the thermal conductivity arising from anharmonicity \cite{lv2016}.
We modified the LAMMPS custom compute command used for GKMA \cite{seyf2019} to incorporate our heat current definition given in Eq.~\eqref{eq:9}, accommodate the centroid-based stress tensor, and enable compatibility with eigenvectors generated by phonopy \cite{togo2015}. This updated implementation was then used for all GKMA calculations in this work. Lastly, we also calculate the phonon density of states (DOS) from MD, which is extracted from the same NVE MD trajectories used for the thermal conductivity calculations by taking the Fourier transform of the velocity auto-correlation function.

\subsection{Phonon quasiparticle method}
Within the phonon quasiparticle approach by Zhang and Wentzcovitch \cite{zhang2021}, which is based on the phonon BTE and the RTA, the lattice thermal conductivity is given as
\begin{equation}
\kappa = \frac{1}{3} \sum_{\mathbf{q}s} C_{\mathbf{q}s} v_{\mathbf{q}s} l_{\mathbf{q}s},
\end{equation}
where $C_{\mathbf{q}s}$, $v_{\mathbf{q}s}$, and $l_{\mathbf{q}s}$ are the mode heat capacity, phonon group velocity, and phonon mean free path (MFP) of mode $(\mathbf{q},s)$, respectively. The MFP is given by
$l_{\mathbf{q}s}=v_{\mathbf{q}s}\tau_{\mathbf{q}s}$, where $\tau_{\mathbf{q}s}$ is the phonon lifetime. The mode heat capacity is calculated as
\begin{equation}
C_{\mathbf{q}s} = \frac{k_B}{V} \frac{x_{\mathbf{q}s} \exp(x_{\mathbf{q}s})}{\left[\exp(x_{\mathbf{q}s}) - 1\right]^2},
\label{eq:heat_capacity}
\end{equation}
with $x_{\mathbf{q}s} = \hbar\widetilde{\omega}_{\mathbf{q}s}/k_BT$, where $\widetilde{\omega}_{\mathbf{q}s}$ is the renormalized frequency. 
To obtain temperature-dependent anharmonic phonon dispersions and group velocities, $\mathbf{v}_i$, the velocity of atom $i$ from the MD trajectory, is projected onto the harmonic modes determined by DFPT as
\begin{equation}
V_{\mathbf{q}s}(t) = \sum_{i=1}^{N} \sqrt{m_i}\ \mathbf{v}_i(t) \cdot \exp\left(i\mathbf{q}\cdot\mathbf{r}_i\right) \cdot \mathbf{e}_{\mathbf{q}s}^*.
\end{equation}
Then, $\tau_{\mathbf{q}s}$ and $\widetilde{\omega}_{\mathbf{q}s}$ are obtained by fitting the mode-projected velocity auto-correlation function to
\begin{equation}
\left\langle V_{\mathbf{q}s}(t_0+t) \cdot V_{\mathbf{q}s}(t_0) \right\rangle = A_{\mathbf{q}s} \cos\left(\widetilde{\omega}_{\mathbf{q}s}t\right) \exp\left(-\frac{t}{2\tau_{\mathbf{q}s}}\right),
\end{equation}
where $A_{\mathbf{q}s}$ is the amplitude. Lastly, the phonon group velocity can be computed as $v_{\mathbf{q}s} = d\widetilde{\omega}_{\mathbf{q}s}/d\mathbf{q}$.

We use Quantum ESPRESSO \cite{giannozzi2009} to compute harmonic phonon frequencies and normal modes via DFPT calculations on a $5 \times 5 \times 5$ q-point mesh in primitive cells with 12 atoms, using the PBEsol XC functional. To generate the MD trajectories, we use the same PBEsol-based MLP employed for the MD simulations in the Green-Kubo method to ensure a direct comparison. NVT simulations are carried out using $5\times5\times5$ supercells containing 1500 atoms for both the seifertite and pyrite-type phases, with a duration of 50 ps and a 1 fs timestep. Temperature-dependent anharmonic phonon dispersions and phonon lifetimes are subsequently extracted from phonon quasiparticles sampled in these MD simulations using the phq \cite{zhang2019} code, with a dense $20 \times 20 \times 20$ q-point mesh.

\section{Results and Discussion}
\subsection{Convergence of thermal conductivity}\label{sec:convergence}
\begin{figure}[t]
\centering
    \includegraphics[width=0.47\textwidth]{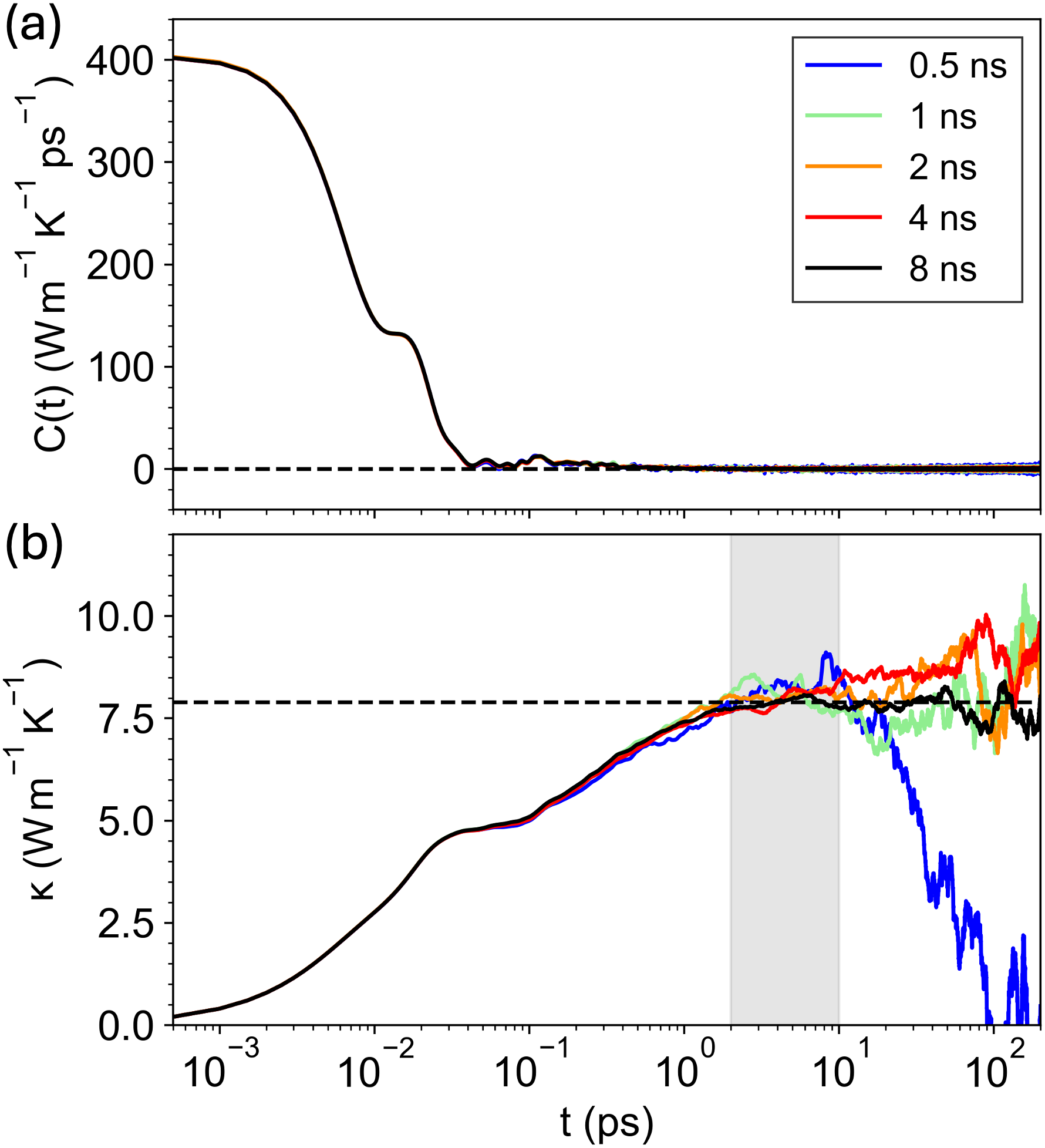}
\caption{\label{fig:1}Heat current auto-correlation function (a) and the corresponding thermal conductivity (b), computed from molecular dynamics (MD) trajectories of different lengths for pyrite-type SiO$_2$ at 200 GPa and 5000 K. The dashed line in panel (b) shows the average value calculated over the gray shaded region using the 8 ns trajectory.}
\end{figure}

To ensure accurate calculations of thermal conductivity values, we first determine an appropriate MD simulation duration. Multiple MD simulations of pyrite-type SiO$_2$ are performed with durations ranging from 0.5 to 8 ns. The heat current auto-correlation function together with the corresponding integrated thermal conductivity are presented in Fig.~\ref{fig:1}. Regardless of the MD simulation duration, the heat current auto-correlation function approaches zero within the first 1 ps as the correlation time $t$ increases. At longer correlation times, the thermal conductivity becomes increasingly affected by statistical errors, requiring long simulations to achieve sufficient averaging. We find that a 2 ns-long simulation is sufficient to accurately calculate the thermal conductivities of both the seifertite and pyrite-type phases. The average thermal conductivity is determined using a 2–10 ps correlation time window.

We also examine finite-size effect by performing simulations with five different system sizes ranging from 96 to 4116 atoms. As shown in Fig.~S2 of the Supplemental Material \cite{SM}, the thermal conductivity initially increases with system size and then reaches a plateau, indicating that a 2592-atom system is sufficient to accurately calculate the lattice thermal conductivity of seifertite and pyrite-type SiO$_2$. We therefore report results obtained from simulations with 2592 atoms.

\begin{figure*}[t]
\centering
    \includegraphics[width=0.84\textwidth]{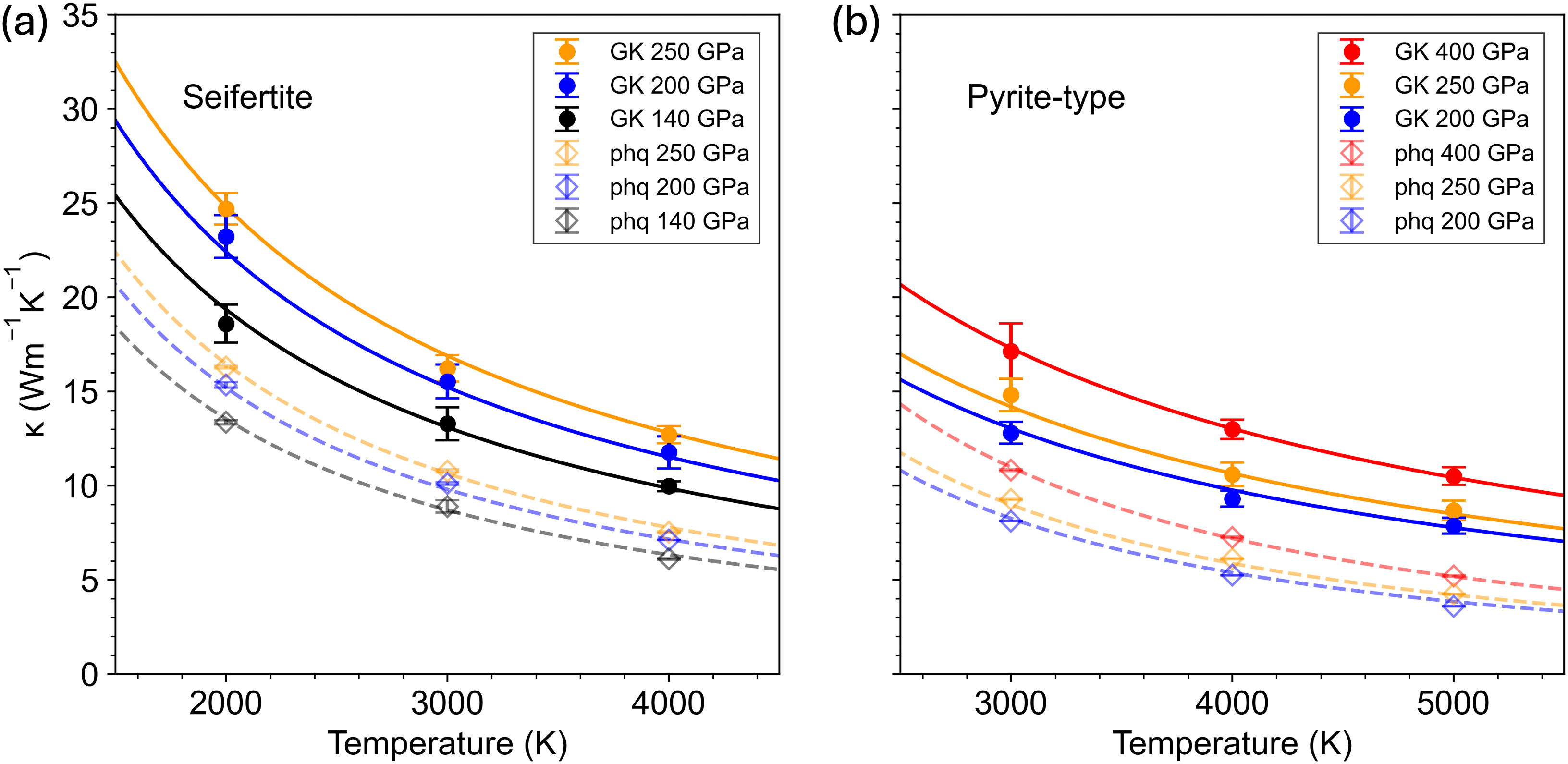}
\caption{\label{fig:2}Thermal conductivity of seifertite (a) and pyrite-type SiO$_2$ (b), calculated using the PBEsol exchange-correlation (XC) functional. Results from the Green-Kubo (GK) method are shown as filled circles with a solid fitted curve, whereas results from the phonon quasiparticle (phq) method are represented by unfilled diamonds with a dotted fitting curve.}
\end{figure*}

\begin{figure}[b]
\centering
    \includegraphics[width=0.45\textwidth]{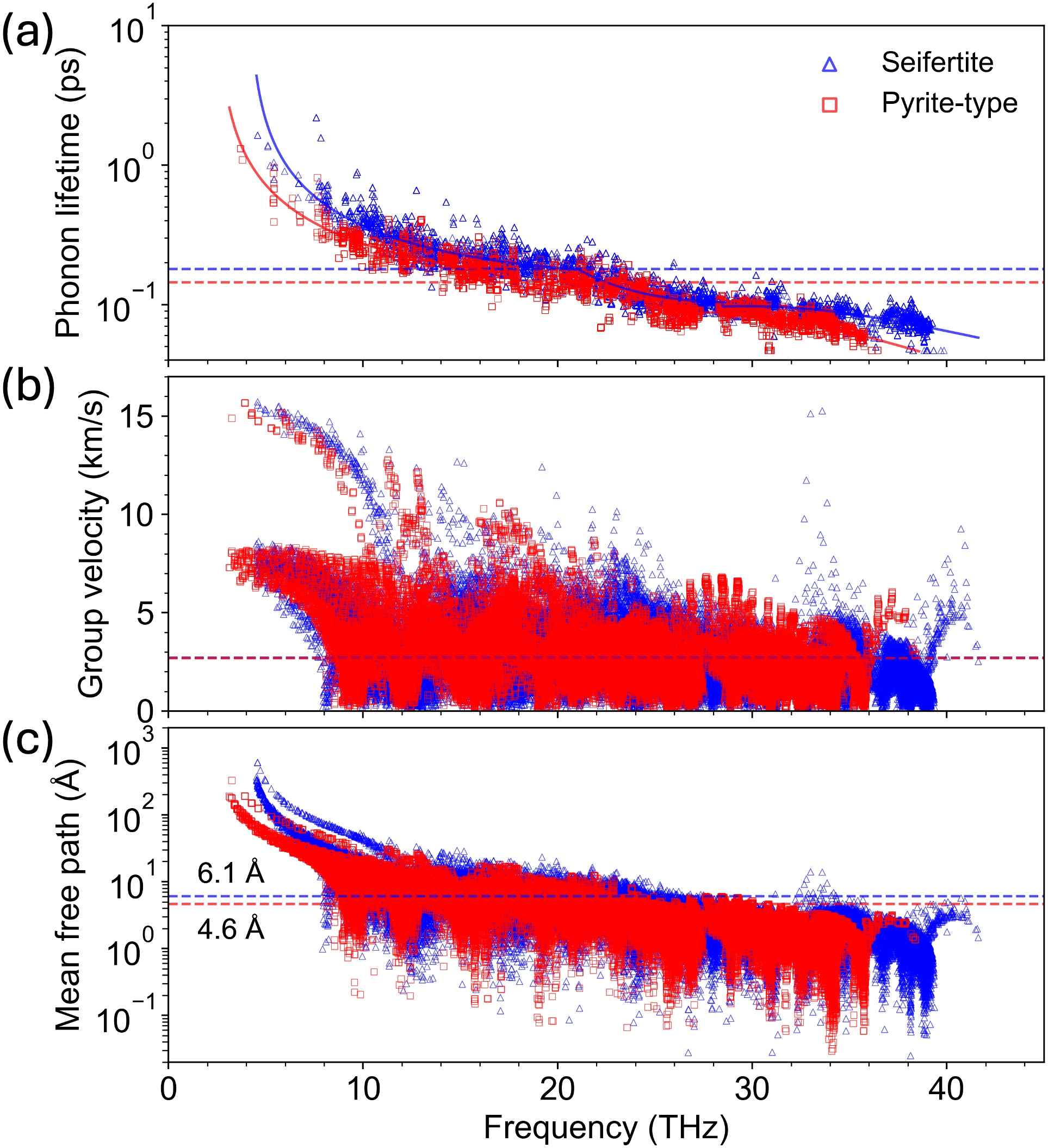}
\caption{\label{fig:3}Phonon lifetimes (a), group velocities (b), and mean free paths (c) for seifertite and pyrite-type SiO$_2$ at 200 GPa and 4000 K, obtained using the phonon quasiparticle (phq) method. Dotted horizontal lines indicate the average values. The solid lines in (a) show the quadratic fits with three segments for each phase.}
\end{figure}

\begin{figure*}[t]
\centering
    \includegraphics[width=0.95\textwidth]{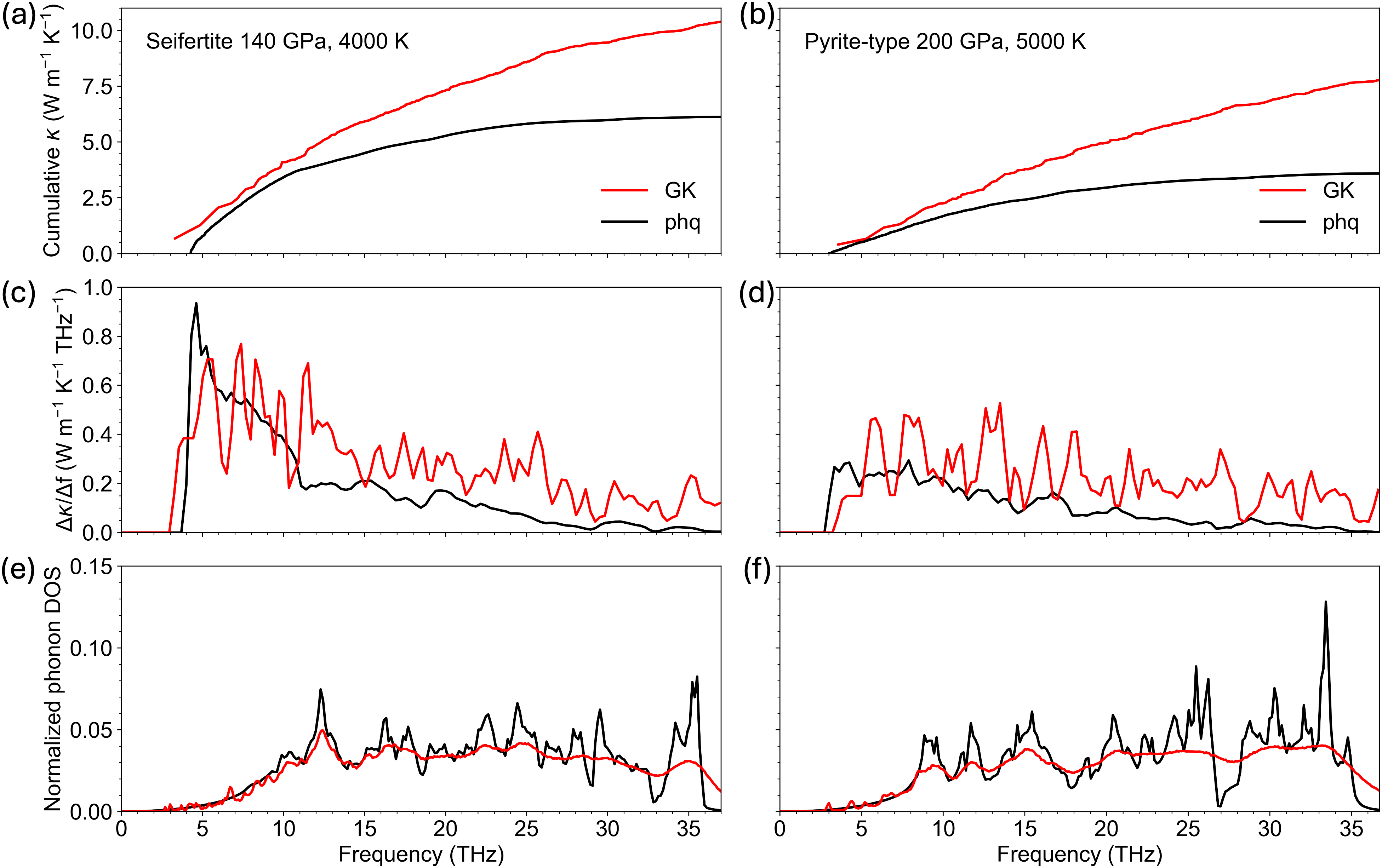}
\caption{\label{fig:4}Comparison of the cumulative thermal conductivity (a),(b), spectral decomposition of thermal conductivity (c),(d), and normalized phonon density of states (DOS) (e),(f) obtained using Green-Kubo (GK) and the phonon quasiparticle (phq) method for seifertite at 140 GPa and 4000 K in (a),(c),(e), and pyrite-type SiO$_2$ at 200 GPa and 5000 K in (b),(d),(f).}
\end{figure*}

\subsection{Thermal conductivity}
Fig.~\ref{fig:2} shows the calculated thermal conductivities of seifertite and pyrite-type SiO$_2$ using the Green-Kubo and phonon quasiparticle methods with the PBEsol XC functional. Our results are well represented by the model \cite{manthilake2011}

\begin{equation}
\label{eq:18}
\kappa = \kappa_{ref} \left( \frac{V_{ref}}{V} \right)^g \left( \frac{T_{ref}}{T} \right)^a,
\end{equation}

with the fitted parameters for each phase and calculation method listed in Table.~S1 of the Supplemental Material \cite{SM}. Reference volumes of 14.8 cm$^3$ mol$^{-1}$ and 14.0 cm$^3$ mol$^{-1}$ are adopted for the seifertite and pyrite-type phases, respectively. A reference temperature $T_{ref} = 3000\,K$ was chosen for both phases. The selection of $V_{ref}$ and $T_{ref}$ does not affect the dependence of thermal conductivity on volume and temperature. The equations of states for both phases are obtained by NPT simulations and are shown in Fig.~S3 and Table.~S2 \cite{SM}.

We first focus on the difference between the Green-Kubo method and the phonon quasiparticle method, both based on the same PBEsol XC functional. In both the seifertite and pyrite-type phases, the phonon quasiparticle method predicts a lower lattice thermal conductivity than the Green-Kubo method. Notably, the discrepancy between the two methods varies by phase, with the pyrite-type phase showing a more pronounced difference. At 200 GPa and 5000 K, which correspond to the lowest pressure and highest temperature conditions among the data points in this work for the pyrite-type phase, the phonon quasiparticle method predicts a thermal conductivity that is 54 \% lower than that obtained from the Green-Kubo method, whereas the maximum difference in seifertite phase is 41 \%. Fig.~\ref{fig:3} shows that, within the phonon quasiparticle method, the pyrite-type phase exhibits significantly shorter average phonon MFP by 25 \% compared to the seifertite phase under the same temperature and pressure conditions of 200 GPa and 4000 K. In Peierls-Boltzmann transport (PBT) theory \cite{sun2010}, on which the phonon quasiparticle method is based, phonons with MFPs below the Ioffe-Regel limit are considered ill-defined \cite{feldman1993}. Although a previous study has shown that the phonon quasiparticle method can identify well-defined phonons even below this limit \cite{zhang2017}, it is inherently unable to capture non-propagating, diffusion-like phonons \cite{simoncelli2022}. We identify the existence of such diffusion-like phonon modes in both seifertite and pyrite-type SiO$_2$ through analyses of eigenvector periodicity and participation ratio (Fig.~S4 of the Supplemental Material \cite{SM}), following the approach of Seyf and Henry \cite{seyf2016}.

To investigate this further, we analyze modal thermal conductivity contributions within the Green-Kubo framework using GKMA and compare the results with those from the phonon quasiparticle method. The GKMA results shown in Fig.~\ref{fig:4} confirm marked differences in modal thermal conductivity contributions across both methods and phases. In seifertite, the GKMA results show that the thermal conductivity contributions are concentrated in the low-frequency modes (red line in Fig.~\ref{fig:4}(c)). In contrast, in the pyrite-type phase, contributions are more broadly distributed across all modes (red line in Fig.~\ref{fig:4}(d)), indicating a noticeable contribution from high-frequency modes. These high-frequency modes have shorter phonon MFPs than low-frequency modes and are therefore more likely to exhibit diffusion-like behavior. Unlike the Green-Kubo method, the phonon quasiparticle method shows highly concentrated thermal conductivity contributions from low-frequency modes in both phases, as indicated by the black lines in Fig.~\ref{fig:4}(c) and (d), with almost negligible contributions from high-frequency modes. Considering all these aspects, the much shorter average MFP obtained for the pyrite-type phase from the phonon quasiparticle method, together with the more strongly weighted contribution of high-frequency modes to the thermal conductivity revealed by GKMA, suggests that the more pronounced discrepancy between the two methods in the pyrite-type phase may originate from diffusion-like, non-propagating phonons with very short MFPs that cannot be adequately captured within the phonon quasiparticle framework.

Furthermore, the phonon DOS (Fig.~\ref{fig:4}(e) and (f)) of pyrite-type exhibits a denser distribution of high-frequency modes, suggesting an increased number of available scattering channels. Possible acoustic-optic hybridization in the pyrite-type phase may enhance scattering \cite{li2016,jia2023}, leading to shorter MFPs that the phonon quasiparticle method cannot fully capture. The difference between the phonon DOS obtained from the quasiparticle-based analysis and that inferred from MD trajectories used in the Green-Kubo method further supports the limitations of the phonon quasiparticle picture at high temperatures. In the quasiparticle description, the vibrational spectrum retains sharp phonon peaks, whereas the MD-derived DOS shows progressive peak broadening and a loss of spectral sharpness with increasing temperature (Fig.~S5 \cite{SM}). This broadening trend is particularly pronounced in the high-frequency region. Such broadening indicates stronger anharmonic damping and reduced phonon lifetimes, suggesting that many vibrational modes, especially high-frequency modes, become less well-defined as independent quasiparticles. Consequently, the phonon quasiparticle framework may underestimate the thermal conductivity when heat transport involves strongly broadened or non-propagating vibrational contributions, which become increasingly significant at elevated temperatures \cite{mukhopadhyay2018,wang2023,peng2024}.

The phonon quasiparticle method predicts the thermal conductivity of the pyrite-type phase at 200 GPa to be 36 \% lower than that obtained using the Green-Kubo method at 3000 K, and the difference increases to 54 \% at 5000 K. The seifertite phase also exhibits the same trend. This increasing difference between the two methods with rising temperature arises from the stronger temperature dependence of the phonon quasiparticle method compared to the Green-Kubo method. For seifertite, the Green-Kubo method yields a temperature dependence of $T^{-0.91}$, whereas the phonon quasiparticle method gives $T^{-1.05}$. Similarly, for the pyrite-type phase, the temperature dependence is $T^{-0.95}$ using the Green-Kubo method and $T^{-1.43}$ using the phonon quasiparticle method. This strong temperature dependence predicted by the phonon quasiparticle method is unlikely to be accurate, considering that nearly all experimental measurements and theoretical calculations for mantle silicates and oxides exhibit temperature dependencies weaker than $T^{-1}$ \cite{xu2004,hunt2011,deschamps2019,wang2022}, except in a study employing the phonon quasiparticle method \cite{zhang2021}. This stronger temperature dependence also can be attributed to the limitations of the phonon quasiparticle method at high temperatures.

\begin{figure}[t]
\centering
    \includegraphics[width=0.48\textwidth]{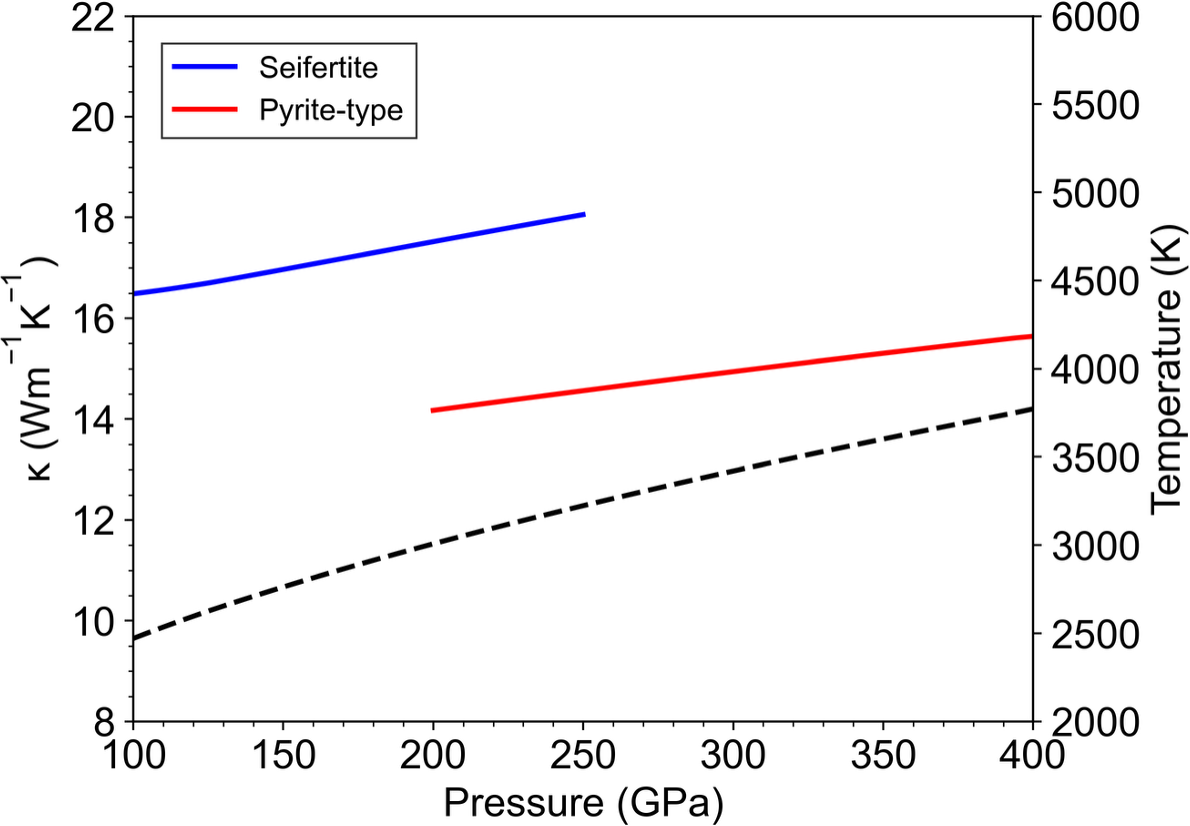}
\caption{\label{fig:5}Thermal conductivities of seifertite (blue line) and pyrite-type SiO$_2$ (red line) along the geotherm (black dashed line) of a super-Earth exoplanet with a core mass fraction of 0.33 \cite{zheng2025}, calculated using the Green-Kubo method with the SCAN-based machine learning potential (MLP).}
\end{figure}

Although the difference between the XC functionals, SCAN and PBEsol (Fig.~S6 of the Supplemental Material \cite{SM}), is less pronounced than the difference between the two methods, it still introduces a non-negligible discrepancy. SCAN consistently predicts higher thermal conductivity than PBEsol, by 9–23 \% for seifertite and 7–14 \% for pyrite-type SiO$_2$. This may arise from the stronger interatomic bonding predicted by SCAN, which is consistent with its prediction of higher melting temperatures and higher densities for these phases compared to PBEsol \cite{park2026}, thereby leading to faster and longer-lived phonons and reduced anharmonicity. Previous benchmarking of lattice thermal conductivity calculations \cite{wei2024} showed that PBEsol provides the best overall accuracy when harmonic phonons are combined with three- and four-phonon scattering, whereas SCAN shows moderate overall performance. However, the same study also indicated that the functional dependence is material-specific. In particular, for MgO, an oxide with partially covalent bonding, SCAN gives more accurate results than the overall trend would suggest. This is relevant for SiO$_2$, where the Si-O bond also has substantial covalent character, and suggests that SCAN may provide a physically meaningful comparison or upper-bound estimate for the thermal conductivity.

Thermal conductivities of seifertite and pyrite-type SiO$_2$ along the geotherm of a super-Earth exoplanet mantle with a mass of 5 M$_{\mathrm{E}}$ \cite{zheng2025} are shown in Fig.~\ref{fig:5}. Across the phase transition from seifertite to the pyrite-type phase at 200–250 GPa and 3000–3200 K, the lattice thermal conductivity decreases by 19 \%. This level of reduction in lattice thermal conductivity across the phase transition is large enough to locally impede conductive heat transport \cite{deschamps2019,guerrero2023}. If the pyrite-type phase is laterally extensive and stable over a finite depth range, it could contribute to the formation of a thermally resistive layer in the deep mantles of super-Earths \cite{naliboff2006}, which may influence mantle convection, core cooling \cite{christensen1995}, and magnetic field generation \cite{christensen2018}, thereby affecting long-term habitability potential \cite{vanhoolst2019}.

\section{Conclusion}
We have computed the thermal conductivities of seifertite and pyrite-type SiO$_2$ by combining the Green-Kubo formalism with \textit{ab initio}-level accuracy MD simulations driven by the MLPs. The Green-Kubo method predicts thermal conductivity values up to 119 \% higher than those obtained from the phonon quasiparticle approach and exhibits weaker temperature dependence for both phases, close to but less than $T^{-1}$, whereas the phonon quasiparticle approach shows a much stronger temperature dependence of up to $T^{-1.42}$. The GKMA results show that the pyrite-type phase has a much more pronounced contribution from high-frequency modes to the thermal conductivity compared to the seifertite phase.

Comparison with results from the phonon quasiparticle approach indicates that the Green-Kubo method captures additional heat transport channels not represented within the quasiparticle framework at high temperatures, where non-propagting, diffusion-like phonons become prevalent but cannot be captured by the phonon quasiparticle approach. The use of the SCAN-based MLP with the Green-Kubo method yields thermal conductivity predictions that are 7–23 \% higher than those obtained using PBEsol, implying stiffer interatomic bonding and reduced anharmonicity. Our results suggest that the phase transition from seifertite to the pyrite-type phase results in a 19 \% decrease in thermal conductivity. This reduction in thermal conductivity across the phase transition may lead to the formation of a thermally insulating layer in the mantles of super-Earths.

\begin{acknowledgments}
We thank D. Zheng for helpful discussions. J.D. acknowledges support from the National Science Foundation (Grant EAR-2444522). All computations for this work were carried out on the Tiger cluster managed and supported by Princeton University's Research Computing. 
\end{acknowledgments}

\section*{Data Availability}
The training and test datasets used to construct and validate the machine learning potential employed in this study are available online via the Open Science Framework \cite{osf_y9qnm}.

\bibliography{refs}
\end{document}

\title{Thermal conductivity of seifertite and pyrite-type SiO\textsubscript{2}: A comparative study}% 

\author{Doyoon Park\,\orcidlink{0009-0007-1224-6525}}
\email{doyoon.park@princeton.edu}
\affiliation{\mbox{Department of Mechanical and Aerospace Engineering, Princeton University, Princeton, New Jersey 08544, USA}}

\author{Yihang Peng\,\orcidlink{0000-0003-4404-8973}}
\affiliation{Department of Geosciences, Princeton University, Princeton, New Jersey 08544, USA}

\author{Jie Deng\,\orcidlink{0000-0001-5441-2797}}
\email{jie.deng@princeton.edu}
\affiliation{Department of Geosciences, Princeton University, Princeton, New Jersey 08544, USA}

\date{\today}
\maketitle

\renewcommand{\thefigure}{S\arabic{figure}}

\begin{figure*}[h]
\centering
    \includegraphics[width=1\textwidth]{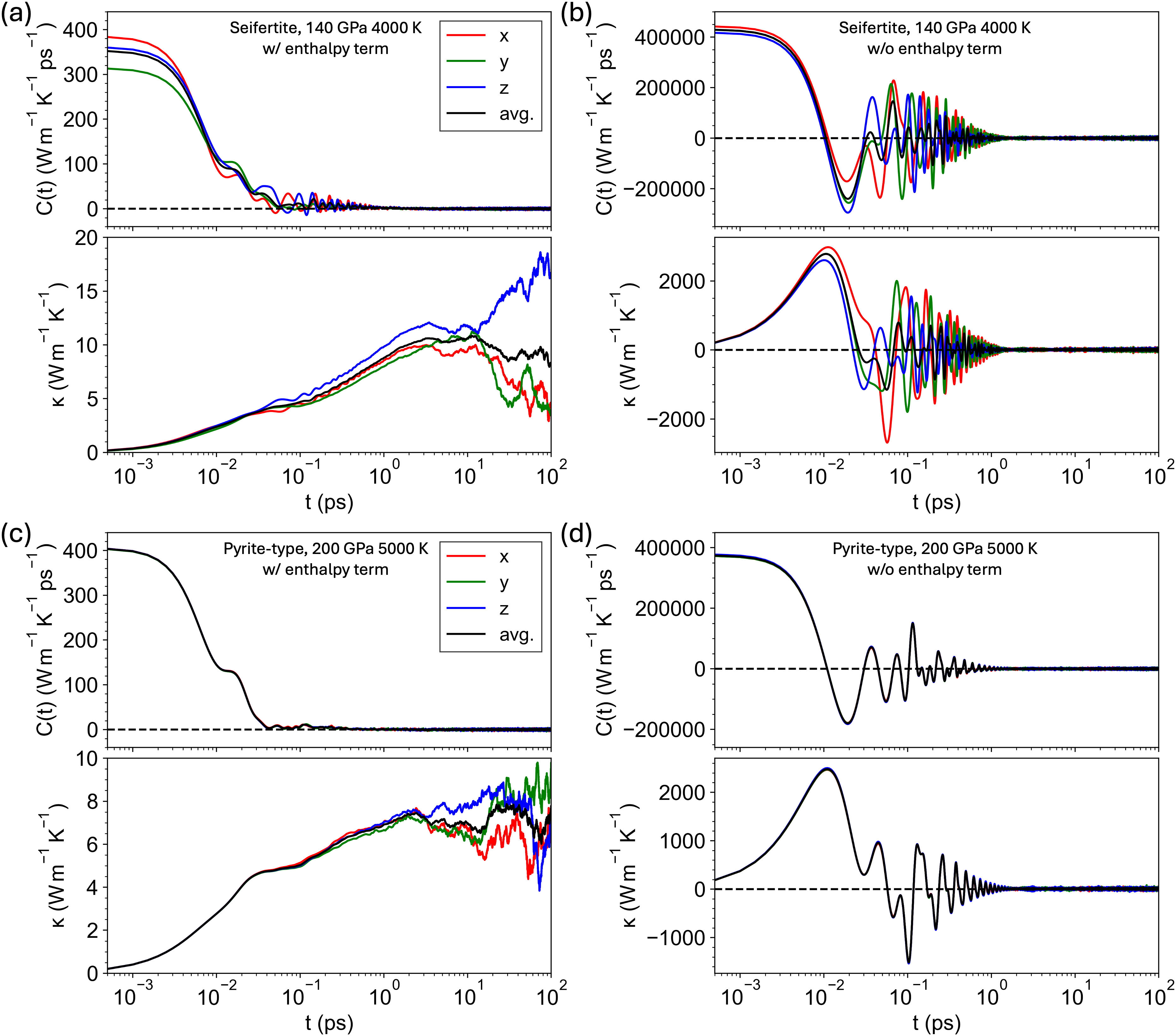}
\caption{\label{fig:s1}Comparison of the heat current auto-correlation function $C(t)$ and the resulting thermal conductivity using different definitions of the heat current, with (a, c) and without (b, d)  subtraction of average partial enthalpy term.}
\end{figure*}

\begin{figure*}[t]
\centering
    \includegraphics[width=0.45\textwidth]{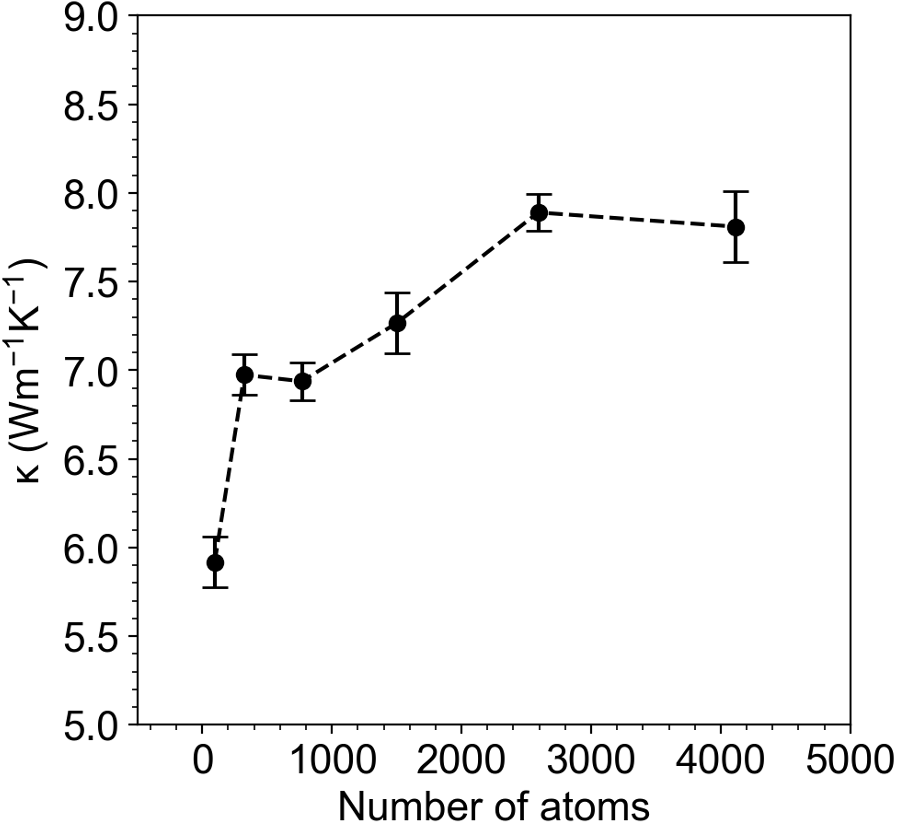}
\caption{\label{fig:s2}Thermal conductivity of pyrite-type SiO$_2$ at 200 GPa and 5000 K for different system sizes. The numbers of atoms in the simulation cells are 96, 324, 768, 1500, 2592, and 4116.}
\end{figure*}

\begin{figure*}[t]
\centering
    \includegraphics[width=0.95\textwidth]{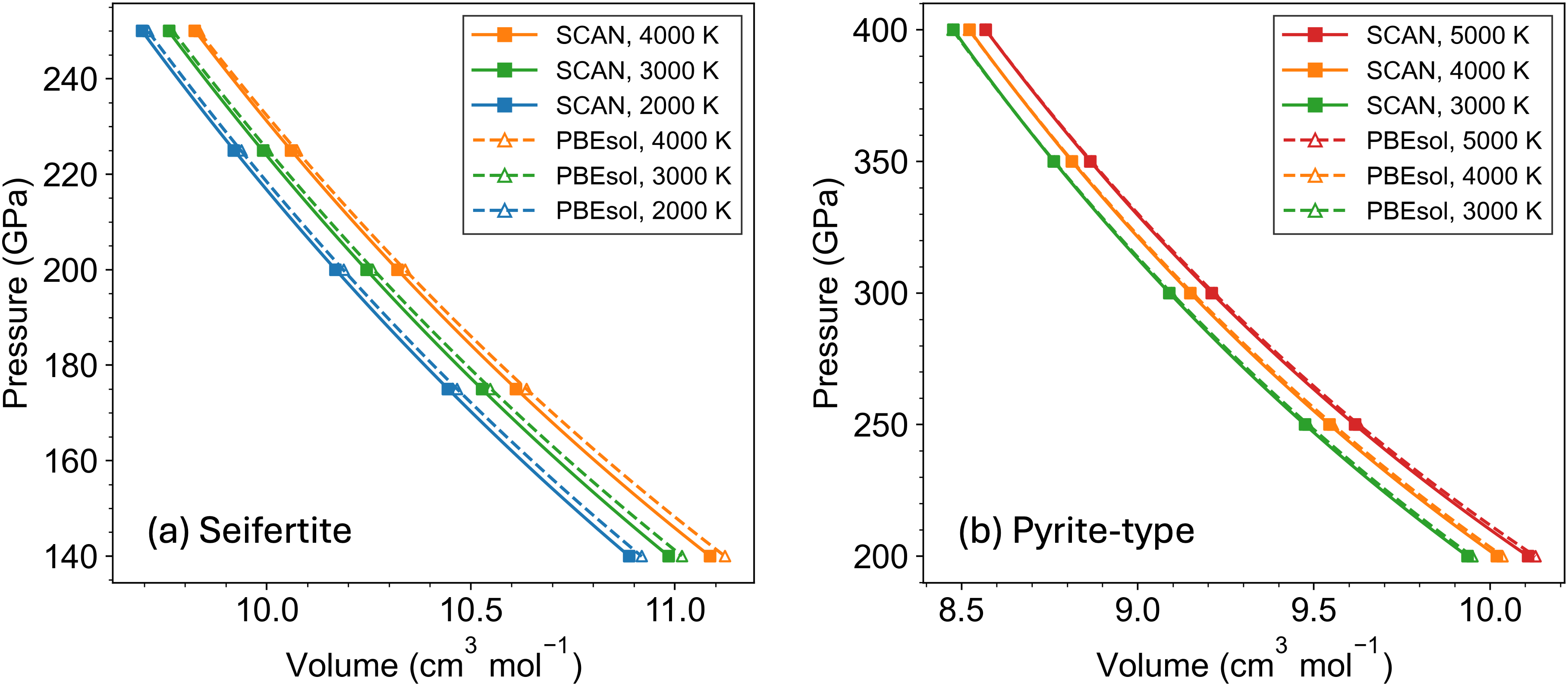}
\caption{\label{fig:s3}Pressure of seifertite and pyrite-type SiO$_2$ as a function of volume at various temperatures. The data points were obtained from molecular dynamics (MD) simulations driven by two different machine learning potentials (MLPs), based on the SCAN (squares with solid lines) and PBEsol (unfilled triangles with dashed lines) exchange-correlation (XC) functionals. The fitted curves represent least squares fits to $P(V, T) = P(V, T_0) + B_{TH}(T-T_0)$, where $P(V, T_0)$ is the reference isotherm at $T_0 = 3000 K$, described by a third-order Birch-Murnaghan equation of state. The thermal pressure coefficient is volume-dependent and given by $B_{TH}(V) = \left[a-b(\frac{V}{V_0})+c(\frac{V}{V_0})^2\right]/1000$, where $a$, $b$, and $c$ are fitting constants. The best fitting parameters are listed in Table~S2.}
\end{figure*}

\setcounter{table}{0}
\renewcommand{\thetable}{S\arabic{table}}

\begin{table}[t]
\caption{Fitted parameters for the thermal conductivity model in Eq.~17 for seifertite and pyrite-type SiO$_2$, obtained using the Green–Kubo (GK) and phonon quasiparticle (phq) methods, as well as two different exchange-correlation (XC) functionals in the GK method.}
\label{tab:s1}
\begin{ruledtabular}
\begin{tabular}{lccccc}
Phase & Method & XC functional & $\kappa_{ref}$ (W m$^{-1}$ K$^{-1}$) & g & a \\
\hline
Seifertite
& GK & SCAN & $8.17\pm0.75$ & $2.07\pm0.24$ & $0.97\pm0.04$ \\
& GK & PBEsol & $6.99\pm0.62$ & $2.12\pm0.24$ & $0.91\pm0.04$ \\
& phq & PBEsol & $5.31\pm0.30$ & $1.67\pm0.15$ & $1.05\pm0.03$ \\
\hline
Pyrite-type
& GK & SCAN & $7.13\pm0.40$ & $2.00\pm0.13$ & $0.93\pm0.04$ \\
& GK & PBEsol & $7.13\pm0.44$ & $1.76\pm0.14$ & $0.95\pm0.05$ \\
& phq & PBEsol & $4.50\pm0.28$ & $1.77\pm0.14$ & $1.43\pm0.05$ \\
\end{tabular}
\end{ruledtabular}
\end{table}

\begin{table}[b]
\caption{Equation of state fitting parameters for seifertite and pyrite-type SiO$_2$ determined from molecular dynamics (MD) simulations driven by two different machine learning potentials (MLPs), based on the SCAN and PBEsol exchange-correlation (XC) functionals.}
\label{tab:s2}
\begin{ruledtabular}
\begin{tabular}{lccccccc}
Phase & XC functional & $V_0$ (cm$^3$ mol$^{-1}$) & $K_0$ (GPa) & $K_0'$ & $a$ & $b$ & $c$ \\
\hline
Seifertite 
& SCAN & $14.58\pm0.05$ & $283\pm7$ & $3.91\pm0.04$ & 0.052 & 0.122 & 0.082 \\
& PBEsol & $14.82\pm0.10$ & $262\pm13$ & $3.97\pm0.08$ & 0.009 & 0.007 & 0.005 \\
\hline
Pyrite-type 
& SCAN & $14.05\pm0.04$ & $291\pm5$ & $3.96\pm0.02$ & 0.015 & 0.020 & 0.013 \\
& PBEsol & $14.02\pm0.03$ & $300\pm4$ & $3.87\pm0.02$ & 0.044 & 0.107 & 0.080 \\
\end{tabular}
\end{ruledtabular}
\end{table}

\begin{figure*}[t]
\centering
    \includegraphics[width=1\textwidth]{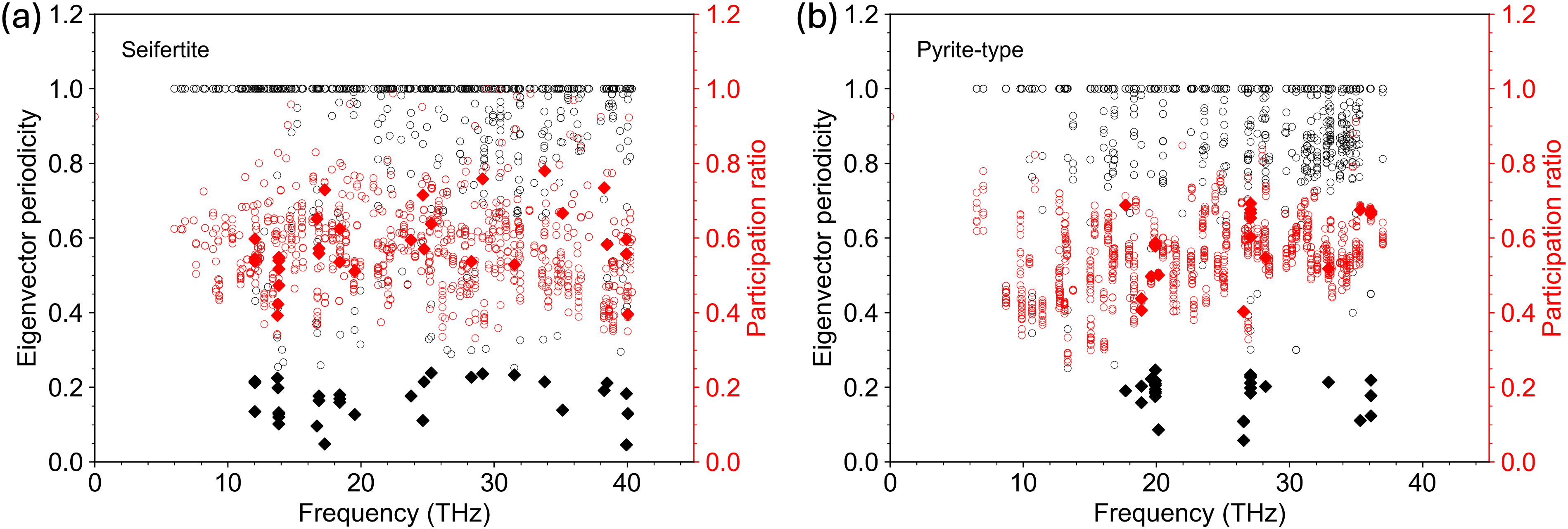}
\caption{\label{fig:s4}Eigenvector periodicity (EP) and participation ratio (PR) analyses, following the approach proposed by Seyf and Henry \cite{seyf2016}, for seifertite (a) and pyrite-type SiO$_2$ (b) at 200 GPa and 4000 K. Diffusion-like modes ($EP < 0.25$ and $PR > 0.1$) are marked by filled diamonds, while propagating modes ($EP > 0.25$ and and $PR > 0.1$) are marked by unfilled circles.}
\end{figure*}

\begin{figure*}[t]
\centering
    \includegraphics[width=0.95\textwidth]{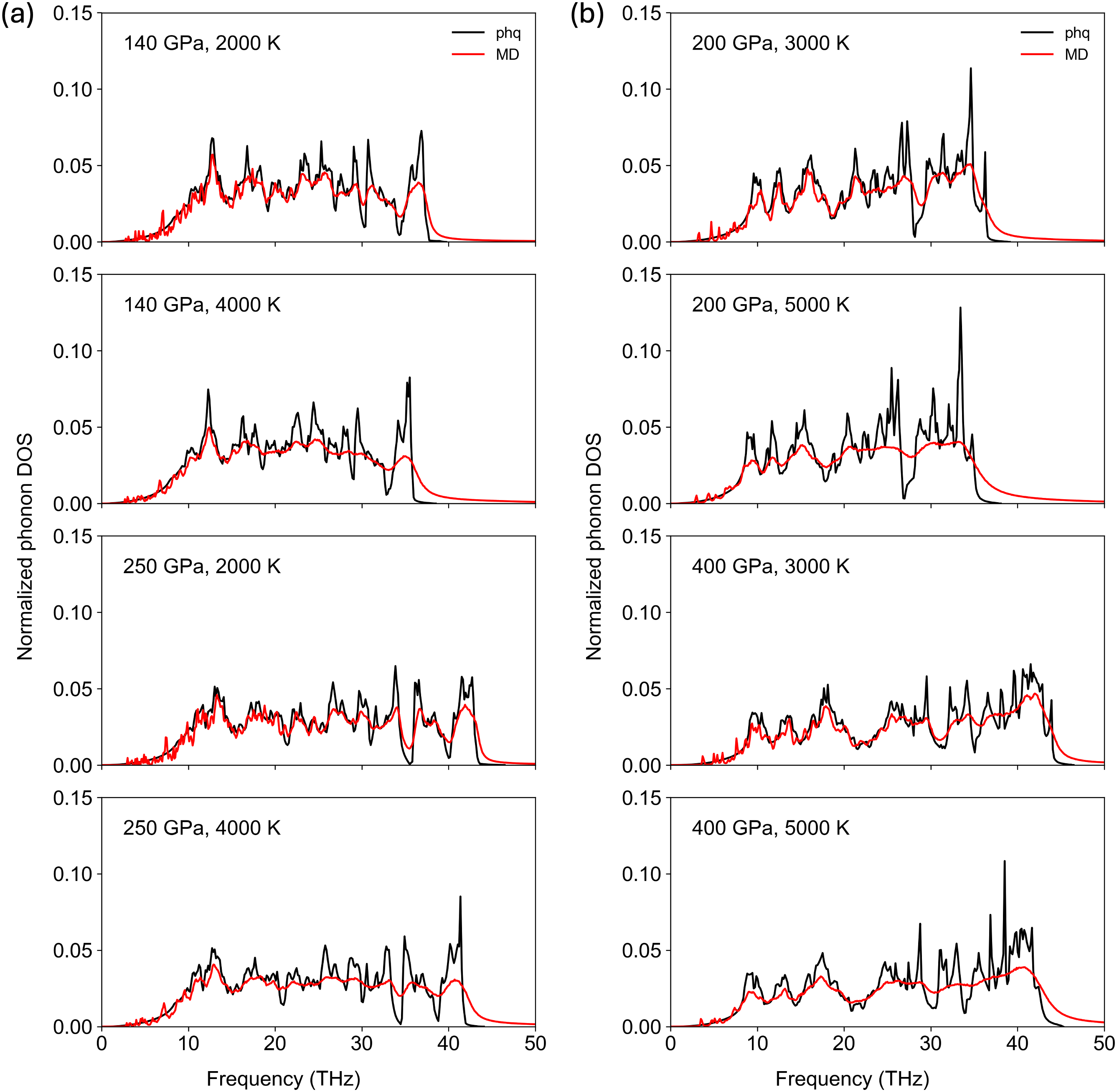}
\caption{\label{fig:s5}Normalized phonon density of states (DOS) for seifertite (a) and pyrite-type (b). Black lines represent the DOS obtained using the phonon quasiparticle (phq) method, while red lines represent the DOS calculated from the Fourier transformation of the velocity auto-correlation function (VACF) from the molecular dynamics (MD) trajectory used in the Green-Kubo method.}
\end{figure*}

\begin{figure*}[t]
\centering
    \includegraphics[width=0.90\textwidth]{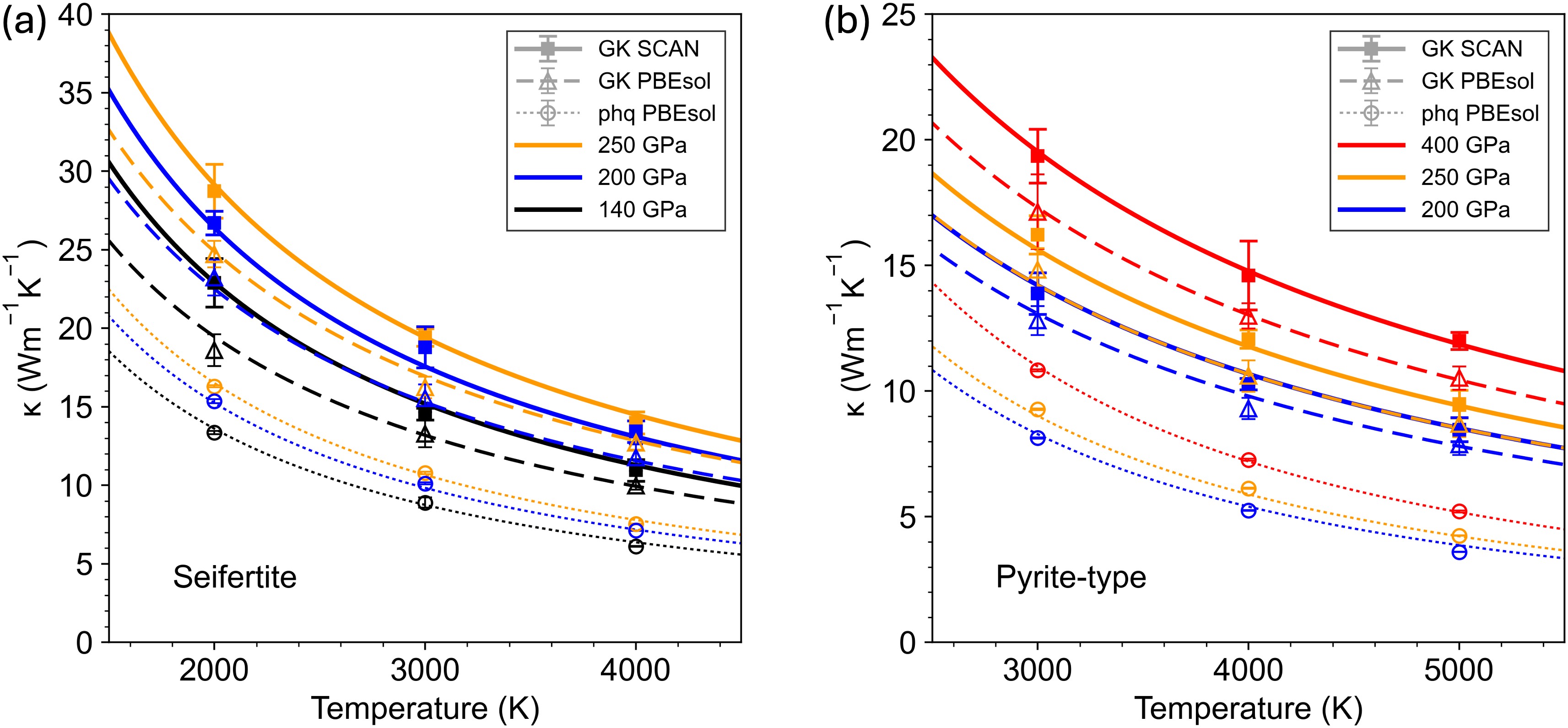}
\caption{\label{fig:s6}Thermal conductivity of seifertite (a) and pyrite-type SiO$_2$ (b). Results from the Green-Kubo (GK) method using the SCAN-based machine learning potential (MLP) are shown as filled squares with a solid fitting curve, while the PBEsol-based results from the GK and phonon quasiparticle (phq) methods are shown as unfilled triangles and circles with dashed and dotted fitting curves, respectively. The colors of the markers and lines indicate the pressure conditions.}
\end{figure*}

\clearpage
\bibliography{refs}